\documentclass[%
 reprint,
superscriptaddress,
 amsmath,amssymb,
 aps,
 prd,
]{revtex4-2}

\usepackage{graphicx}% Include figure files
\usepackage{dcolumn}% Align table columns on decimal point
\usepackage{bm}% bold math
\usepackage{xcolor}
\usepackage{mathtools}
\usepackage{titlesec}
\usepackage{svg}
\usepackage{hyperref}
\usepackage{amsmath,amsthm}
\usepackage{braket}
\usepackage{pifont}
\usepackage{enumitem}
\renewcommand{\qed}{\hfill \square}
\newtheorem{theorem}{Theorem}%[section]
\newtheorem{corollary}{Corollary}[theorem]
\newtheorem{lemma}[theorem]{Lemma}

\begin{document}

% \preprint{APS/123-QED}

\title{Stabilizer Tensor Networks: universal quantum simulator on a basis of stabilizer states}

\author{Sergi Masot-Llima}
\affiliation{Barcelona Supercomputing Center, Barcelona 08034, Spain}
\affiliation{Universitat de Barcelona, Barcelona 08007, Spain}

\author{Artur Garcia-Saez}
\affiliation{Barcelona Supercomputing Center, Barcelona 08034, Spain}
\affiliation{Qilimanjaro Quantum Tech, Barcelona 08007, Spain}

\date{\today}

\begin{abstract}
    Efficient simulation of quantum computers relies on understanding and exploiting the properties of quantum states. This is the case for methods such as tensor networks, based on entanglement, and the tableau formalism, which represents stabilizer states. In this work, we integrate these two approaches to present a generalization of the tableau formalism used for Clifford circuit simulation. We explicitly prove how to update our formalism with Clifford gates, non-Clifford gates, and measurements, enabling universal circuit simulation. We also discuss how the framework allows for efficient simulation of more states, raising some interesting questions on the representation power of tensor networks and the quantum properties of resources such as entanglement and magic, and support our claims with simulations.
\end{abstract}

%\keywords{Quantum computing -- Quantum simulation -- Tensor Networks}

\maketitle

% \tableofcontents

Simulation of quantum computing is crucial for two main reasons: driving science in fields like condensed matter physics \cite{vidalClassQuantumManyBody2008,frerotProbingQuantumCorrelations2023a} or quantum chemistry \cite{shangLargescaleSimulationQuantum2022a,daltonQuantifyingEffectGate2024a,dalzellQuantumAlgorithmsSurvey2023}, as long as we do not have large, error-corrected devices, and testing quantum advantage claims \cite{liuValidatingQuantumsupremacyExperiments2022,tindallEfficientTensorNetwork2024,begusicFastConvergedClassical2024} made by cutting-edge devices \cite{aruteQuantumSupremacyUsing2019,kimEvidenceUtilityQuantum2023a}. To simulate efficiently beyond a few dozen qubits, we must find alternative characterisations due to the exponential growth of brute force approaches. Thus, a large effort is put towards identifying which states are easy and why, given the absence of a universal description of simulablity. Resource theories \cite{RevModPhys.91.025001} are a useful tool for this task: they characterize the operations that are easy to do (free operations) in a certain framework. We are particularly interested in entanglement \cite{christandlResourceTheoryTensor2023} and stabilizer rank \cite{bravyiSimulationQuantumCircuits2019b,veitchResourceTheoryStabilizer2014}, for their relation to tensor networks (TN) and the stabilizer formalism, respectively.\\

\begin{figure*}[!ht]
\centering
\includegraphics[scale=0.169]{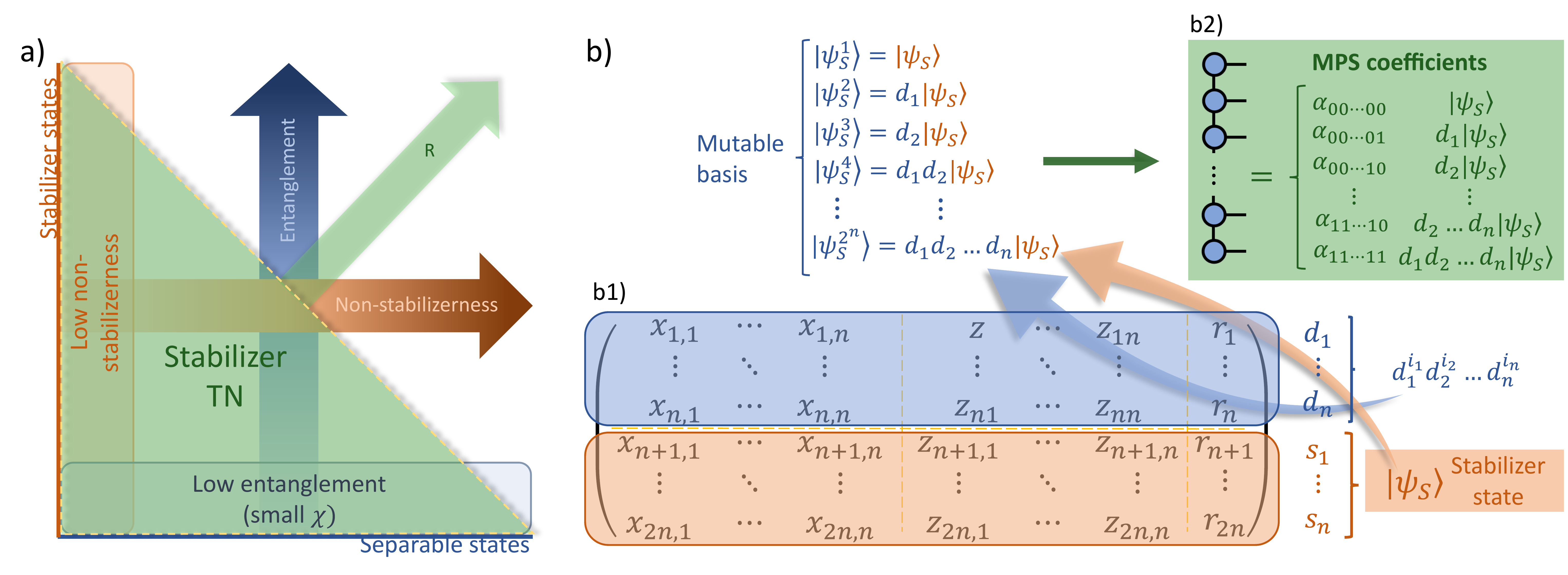}\caption{Showcase of the stabilizer formalism. In a), we classify states under two different resources. For the resource of entanglement, in blue, (non-stabilizerness, orange), axis y=0 (x=0) represents the free states, while its adjacent region represents states with low amounts of entanglement (non-stabilizerness), which are classically simulatable with tensor networks (stabilizer tableaus). They are simultaneously simulatable with stabilizer tensor networks, in green, and can be characterized with a different resource $R$. In b) we show how stabilizer TN joins the other methods: the tableau formalism (b1) encodes a stabilizer state and a set of destabilizer generators which are used to form a basis $\mathcal{B}(\mathcal{S},\mathcal{D})$ for the Hilbert space. The first $n$ rows of the tableau encode the decomposition of $n$ destabilizer generators, and rows $n+1$, ... , $2n$ encode $n$ stabilizer generators. An extra column $r$ indicates the phase of each generator, since $S$ and $-S$ stabilize different states. The amplitudes of decomposing a state $\ket{\psi}$ into $\mathcal{B}(\mathcal{S},\mathcal{D})$ are stored in a tensor network (green, b2).}
\label{fig:main}
\end{figure*}

The interest in relating different resources, particularly these two, is not new. Previous research has found some of these states present maximal entanglement, at least in the bipartite sense \cite{smithTypicalEntanglementStabilizer2006}, although most types of entangled states are not achievable with these circuits. Recently, magic in Matrix Product States (MPS), a type of TN, has also been characterized and looked into \cite{haugQuantifyingNonstabilizernessMatrix2022,tarabungaNonstabilizernessMatrixProduct2024,lamiLearningStabilizerGroup2024}, and it is noteworthy that separable states with a lot of magic are complex in the stabilizer formalism, even though they are trivial to simulate with resource theories of entanglement. This means that these resources are, in some sense, orthogonal, as depicted in Fig. \ref{fig:main}a. In this article, we unify simulation strategies for entanglement and magic by using a special basis \cite{yoderGeneralizationStabilizerFormalism} in conjunction with tensor networks, as shown in Fig. \ref{fig:main}b, and we focus on how the proposed method can simulate arbitrary circuits.\\

Entanglement as a resource for simulation is usually characterized by bipartite entanglement between sectors of the system. This applies to methods such as circuit cutting \cite{pengSimulatingLargeQuantum2020}, entanglement forging \cite{eddinsDoublingSizeQuantum2022} or tensor networks \cite{orusPracticalIntroductionTensor2014}. These simulations rely on limited entanglement, mostly between close neighbors \cite{ciracMatrixProductStates2021}, or on a hierarchical structure of entanglement \cite{vidalEntanglementRenormalization2007,evenblyTensorNetworkStates2011}. Free operations consist of single-qubit (local) gates and classical communication \cite{bennettQuantumNonlocalityEntanglement1999,chitambarEverythingYouAlways2014}. In the extreme case of no entanglement, the simulation cost of a system grows linearly with its size; in more complex cases, systems can adhere to an area law \cite{eisertColloquiumAreaLaws2010} that allows TN methods like DMRG \cite{mccullochInfiniteSizeDensity2008d} to perform efficient simulations with great success. Tensor networks encode high-dimensional tensors into the product of smaller, low-dimensional ones, and are in general advantageous whenever long-range correlations are restricted. The tensors can be connected through bonds in various geometries, and networks with more complexity and expressive power will entail higher performance costs. We focus on a 1D MPS structure to encode the amplitudes of a quantum state:
\begin{equation}\label{eq:basic_tn}
    \mathcal{T}^{i_1 i_2 \dots i_n} = \sum_{k_1k_2\dots k_{n-1}} (T_1)^{i_1}_{k_1}(T_2)^{i_2 k_1}_{k_2}(T_3)^{i_3 k_2}_{k_3} \dots (T_N)^{i_n k_{n-1}}
\end{equation}
In this structure, the dimension $\chi$ of a given bond $k$ corresponds to the entropy between the two subsystems it connects, as measured by the Schmidt rank \cite{Nielsen_Chuang_2010}. We call this $\chi$ the bond dimension. A separable state can be encoded into a TN with $\chi=1$, whereas a state with mostly local entanglement (AKLT state \cite{affleckRigorousResultsValencebond1987}) needs $\chi=2$, and a maximally entangled state requires up to $\chi=2^{n/2}$. The bond dimension can also be artificially limited at the cost of precision. \\

The stabilizer tableau formalism \cite{gottesmanStabilizerCodesQuantum1997}, on the other hand, can simulate efficiently any circuit composed only by Clifford gates with a classical computer. The states that can be prepared under these constraints are known as stabilizer states. In this context, the set $\mathcal{C}$ of Clifford gates are the free operations, and non-Clifford gates increase the stabilizer rank \cite{pelegLowerBoundsStabilizer2022}, which constitutes the resource. However, resource theories of stabilizer states are typically studied with other measures such as magic \cite{haugStabilizerEntropiesNonstabilizerness2023}, stabilizer Rényi entropy \cite{leoneStabilizerRenyiEntropy2022} or Wigner positivity \cite{mariPositiveWignerFunctions2012a} due to their interesting properties. This simulation formalism is based on a stabilizer set $\mathcal{S}$ of Pauli operators ($\mathcal{P}_n$). It uniquely defines a state that fulfills $S\ket{\psi_{\mathcal{S}}}=\ket{\psi_{\mathcal{S}}}$ for any $S\in\mathcal{S}$. A Pauli operator P can be described with two boolean vectors and a phase, $P = \alpha \cdot (x_1 x_2 \dots x_n) \cdot (z_1 z_2 \dots z_n)$, meaning we only need $2n+1$ boolean values to represent one. Also, a set of $n$ generators of $\mathcal{S}$ are enough to fully define the group. This means a tableau of $n\times(2n+1)$ boolean entries stores all the information about $\mathcal{S}$ (see Fig. \ref{fig:main}, b1). Since $\mathcal{P}_n$ is closed under the action of any gate $C\in\mathcal{C}$, a Clifford circuit can be simulated by finding the new tableau after applying each gate $C$ in it. An efficient ($O(n^2)$ time) approach to update the tableaus is known \cite{aaronsonImprovedSimulationStabilizer2004}, and it works by also storing the generators $d_i$ of the destabilizer group $\mathcal{D}$ -- these operators fulfill $\{d_i,s_i\}=0,[d_i,d_j]=0$ and $[d_i,s_j]=0$ for any $i\neq j$.\\

In the following equations, we use the notation $d_{\hat{i}}$ for a generic destabilizer in $\mathcal{D}$, defined by $\hat{i}$ and the generators $d_i$ as $d_{\hat{i}}=d_1^{i_1}\dots d_n^{i_n}$; the same follows for stabilizers $s_{\hat{i}}$. Paired with the stabilizer state $\ket{\psi_{\mathcal{S}}}$, they define the set $\{d_{\hat{i}}\ket{\psi_{\mathcal{S}}}\}_{\hat{i}}$, which forms a basis $\mathcal{B}(\mathcal{S},\mathcal{D})$\cite{yoderGeneralizationStabilizerFormalism} of the Hilbert space $\mathcal{H}^n$:
\begin{equation}\label{eq:nu}
    \ket{\psi} = \sum_{i=0}^{2^n} \nu_i d_{\hat{i}} \ket{\psi_\mathcal{S}}.
\end{equation}
This is shown and proven in \hyperref[lemma:representation]{Lemma \ref{lemma:representation}}. We encode these amplitudes on an MPS using $\ket{\nu}=\sum_i \nu_i \ket{i}$, and show how they change when applying any unitary gate or measurement with the following \textit{update rules}:\\

\begin{enumerate}[leftmargin=*]
    \item \textbf{Clifford gate $G$}: Update the stabilizer basis $\mathcal{B}(\mathcal{S},\mathcal{D})$ by conjugating with $G$, following the rules in the tableau formalism (see \cite{aaronsonImprovedSimulationStabilizer2004} or Appendix \ref{sec:tableau_rules}) for the update $\ket{\psi_\mathcal{S}} \rightarrow G\ket{\psi_\mathcal{S}}=\ket{\psi_{\tilde{\mathcal{S}}}}$. This gives a new basis $\mathcal{B}(\tilde{\mathcal{S}},\tilde{\mathcal{D}})$:
    \begin{equation}\label{eq:rule1_psi}
        G\ket{\psi} = \sum_i \nu_i G d_i \ket{\psi_{\mathcal{S}}} = \sum_i \nu_i \tilde{d}_i \ket{\psi_{\tilde{\mathcal{S}}}},
    \end{equation}
    and leaves the coefficient state $\ket{\nu}$ unchanged
    \begin{equation}\label{eq:rule1_nu}
        G\ket{\nu} = \ket{\nu}.
    \end{equation}

    \item \textbf{Non-Clifford gate $\mathcal{U}$}: Find the decomposition \\ $\mathcal{U}=\sum_i\phi_i\delta_{\hat{d}_i}\sigma_{\hat{s}_i}$, then modify $\ket{\psi}$ as:
    \begin{equation}\label{eq:rule2_psi}
        \mathcal{U}\ket{\psi} = \sum_{i,j} ((-1)^{j \cdot {\hat{s}_i}} \phi_i \nu_j) \: d_{j+\hat{d}_i} \ket{\psi_{\mathcal{S}}}.
    \end{equation}
    When the decomposition only has two terms, this is equivalent to a rotation on $\ket{\nu}$:
    \begin{equation}\label{eq:rule2_nu}
        \ket{\nu'} = \cos(\theta) I - i \sin(\theta) X_{I_x} Y_{I_y} Z_{I_z}\ket{\nu}.
    \end{equation}
    The basis $\mathcal{B}(\mathcal{S},\mathcal{D})$ stays unchanged.

    \item \textbf{Measurement of observable $O$}: Find the decomposition \\ $O=\alpha \delta_{\hat{d}} \sigma_{\hat{s}}$ and the value of $\braket{\mathcal{O}}$ with:
    \begin{equation}
        \braket{\mathcal{O}} = \alpha \braket{\nu | X_{\hat{d}} Z_{\hat{s}} | \nu}.
    \end{equation}
    Now choose an outcome $m\in\{+,-\}$ with probability $p_+ = \frac{I+\braket{O}}{2}$, $p_- = 1-p_+$, and let $k$ be the position of the first 1 in $\hat{d}$. Then update the stabilizer basis to $\mathcal{B'}(\mathcal{S'},\mathcal{D'})$ following the rules for a measurement in the original formalism, and find $\ket{\psi'}=(I+m\mathcal{O})/2\ket{\psi}$ as:
    \begin{equation}\label{eq:rule3_psi}
        \ket{\psi'} = \sum_i \delta_{i_k,0} \left( \frac{1}{\sqrt{2}} \nu_{\hat{i}} + m\frac{\alpha (-1)^{\hat{i} \cdot \hat{s}}}{\sqrt{2}} \nu_{\hat{i}+\hat{d}} \right) d_{\hat{i}} \ket{\psi_\mathcal{S'}},
    \end{equation}
    which equals to a projection and rotation on $\ket{\nu}$:
    \begin{equation}\label{eq:rule3_nu}
        \ket{\nu'}= \ket{0}\bra{0}_k \left( \frac{1}{\sqrt{2}} I + m \frac{\alpha (-i)^{|I_y|}}{\sqrt{2}} X_{I_x} Y_{I_y} Z_{I_z} \right) \ket{\nu}.
    \end{equation}
\end{enumerate}

When simulating a circuit, we use a decomposition into CNOT and single qubit rotations, as is usually done in real devices. This ensures that all non-Clifford gates conform to the particular case of Eq. \ref{eq:rule2_nu}. The measured observables $O$ are decomposed into (de)stabilizers, which we distinguish from the generators of the basis $d_{\hat{i}}$ by writing $\sigma_{\hat{s}}$ ($\delta_{\hat{d}}$) instead. In \hyperref[sec:lemmas]{Annex \ref{sec:lemmas}} we explain the rules in more detail and prove two more Lemmas that justify the equations shown here. To compute the change of basis for Clifford gates, we employ already known \cite{aaronsonImprovedSimulationStabilizer2004} efficient methods to update the tableau. Since the amplitudes do not change, they preserve $\chi$. Non-Clifford gates and measurements, on the other hand, can introduce correlations between amplitudes that increase $\chi$ and make calculations more expensive. Consequently, $\chi$ constitutes our resource and the free operations include all Clifford gates \textit{and also} non-Clifford gates $\mathcal{U}$ such that Eq. \ref{eq:rule2_nu} is a local rotation on $\ket{\nu}$. We prove this in \hyperref[lemma:free]{corollary \ref{lemma:free}}, in the annex. \\

The key ingredient to stabilizer tensor networks is allowing the basis to change beyond local rotations. The tableau algorithm replaces the computational basis with a basis of stabilizer states, which can have some entanglement, and forgoes the correspondence between qubits and tensors. In a way, entanglement is transferred from the tensor network $\ket{\nu}$ representation into the basis, at the cost of single qubit gates potentially becoming entangling on the amplitudes of $\ket{\nu}$. In general, this only happens if the circuit already contained entangling gates, and thus that part of the circuit was entangling to begin with. Therefore, we argue that the formalism does not generate fictitious entanglement. Instead, we say we store \textit{potential entanglement} in the basis.\\

We mentioned several resources linked to non-stabilizerness. Among those, stabilizer rank has a direct link to our formalism. For an arbitrary state $\ket{\psi}$, its stabilizer rank is the smallest $\xi$ that allows a decomposition into stabilizer states $\ket{\psi_S}$:
\begin{equation}\label{eq:stab_rank}
    \ket{\psi} = \sum_{i=1}^\xi \alpha_i \ket{\psi_S^i}.
\end{equation}
Stabilizer states have $\xi=1$. The structure on the basis states we use does not mean that a low stabilizer rank translates into a simple $\ket{\nu}$, as the necessary stabilizer states might not be simultaneously in $\mathcal{B}(\mathcal{S},\mathcal{D})$. We can define a pseudo-stabilizer rank $\tilde{\xi}$ as the amount of non-zero coefficients in $\ket{\nu}$. This is obviously an upper bound to $\xi$, but a thorough characterization of how these quantities relate is left for future work.\\

Let us demonstrate that our formalism can efficiently simulate two different scenarios: low entanglement and low stabilizer rank. In the original tableau formalism \cite{aaronsonImprovedSimulationStabilizer2004}, it was already shown how we can simulate any circuit and encode any state in the $n$ qubit Hilbert space with a superposition of tableaus. However, this is akin to a brute-force simulation with the statevector approach, and grows exponentially with the number $t$ of non-Clifford gates. Instead, our formalism can take advantage of all the tools that have been developed for tensor networks simulations. Consider the state $\ket{T}^n$, which can be prepared with:
\begin{equation}\label{eq:t-state}
    \ket{T}^n =  \prod_{i=1}^{n} T_i \prod_{i=1}^{n} H_i \ket{0}^{\otimes n}.
\end{equation}
The first layer of Hadamards, which are Clifford gates and only updates the tableau, sets the stabilizer basis to $s_i=X_i,d_i=Z_i$. In this basis, each $T$-gate on qubit $i$
\begin{equation}\label{eq:t_example}
    T_i = \cos(\frac{\pi}{8}) I - i \sin(\frac{\pi}{8}) Z_i = \cos(\frac{\pi}{8}) I - i \sin(\frac{\pi}{8}) d_i,
\end{equation}
fulfills the criteria for a free operation, so the resulting state is represented by a trivial MPS with $\chi=1$. Notice that, in this case, the pseudo-stabilizer rank is maximal $\tilde{\xi}=2^n$. With the conventional generalization of tableaus, we would need $2^n$ copies, as each $T$-gate duplicates the number of necessary tableaus. This is not the best that can be achieved: the stabilizer rank of $\ket{T}^n$ for small $n$ has been shown to be low \cite{qassimImprovedUpperBounds2021a}, meaning an optimal decomposition requires fewer tableaus (and also $\xi <<\tilde{\xi}$). However, a general method to find these decompositions is not known. Most importantly, the growth of $\xi$ with $n$ is expected to be exponential \cite{bravyiTradingClassicalQuantum2016} unless quantum computing is completely simulatable (even though a supralinear lower bound has not been found \cite{pelegLowerBoundsStabilizer2022}), whereas stabilizer tensor networks can represent these states efficiently for any $n$. On the other hand, some stabilizer states have been shown to have maximum bipartite entanglement \cite{smithTypicalEntanglementStabilizer2006}, as mentioned earlier. These states can be prepared with a Clifford circuit, so in a simulation with stabilizer tensor networks they will be an element of the basis $\mathcal{B}(\mathcal{S},\mathcal{D})$, and therefore trivial to represent with $\xi=\tilde{\xi}=1$, despite being expensive with a regular MPS.\\

In addition to these examples, there likely exists a different resource $R$ that captures the power of the approach, defining whether a state can be efficiently represented or not with a single metric as illustrated in Fig. \ref{fig:main}a. This resource $R$ must be related to the two discussed resources, in the sense that both low entanglement or low stabilizer rank imply low $R$, because we have seen that we can simulate either case. However, since stabilizer states can be very entangled, and separable states can have high non-stabilizerness, it follows that high entanglement does not imply high $R$, nor does low stabilizer rank imply high $R$, indicating that it isn't trivially connected to these resources.\\

Beyond the cases of complete stabilizerness or no entanglement, the efficiency should persist when these resources are present in a low amount for the formalism to be useful. Our discussion so far highlights two advantages in that regard. First, notice that we can always process Clifford gates directly on $\ket{\nu}$ instead of changing the basis so that the TN behaves traditionally. This means any state simulatable with tensor networks is also feasible in our approach. Additionally, we have seen that Clifford gates don't change $\ket{\nu}$, independently of its $\chi$, so we can always move freely in the space of states with fixed stabilizer rank.
\begin{figure}[!ht]
\centering
\includegraphics[scale=0.3]{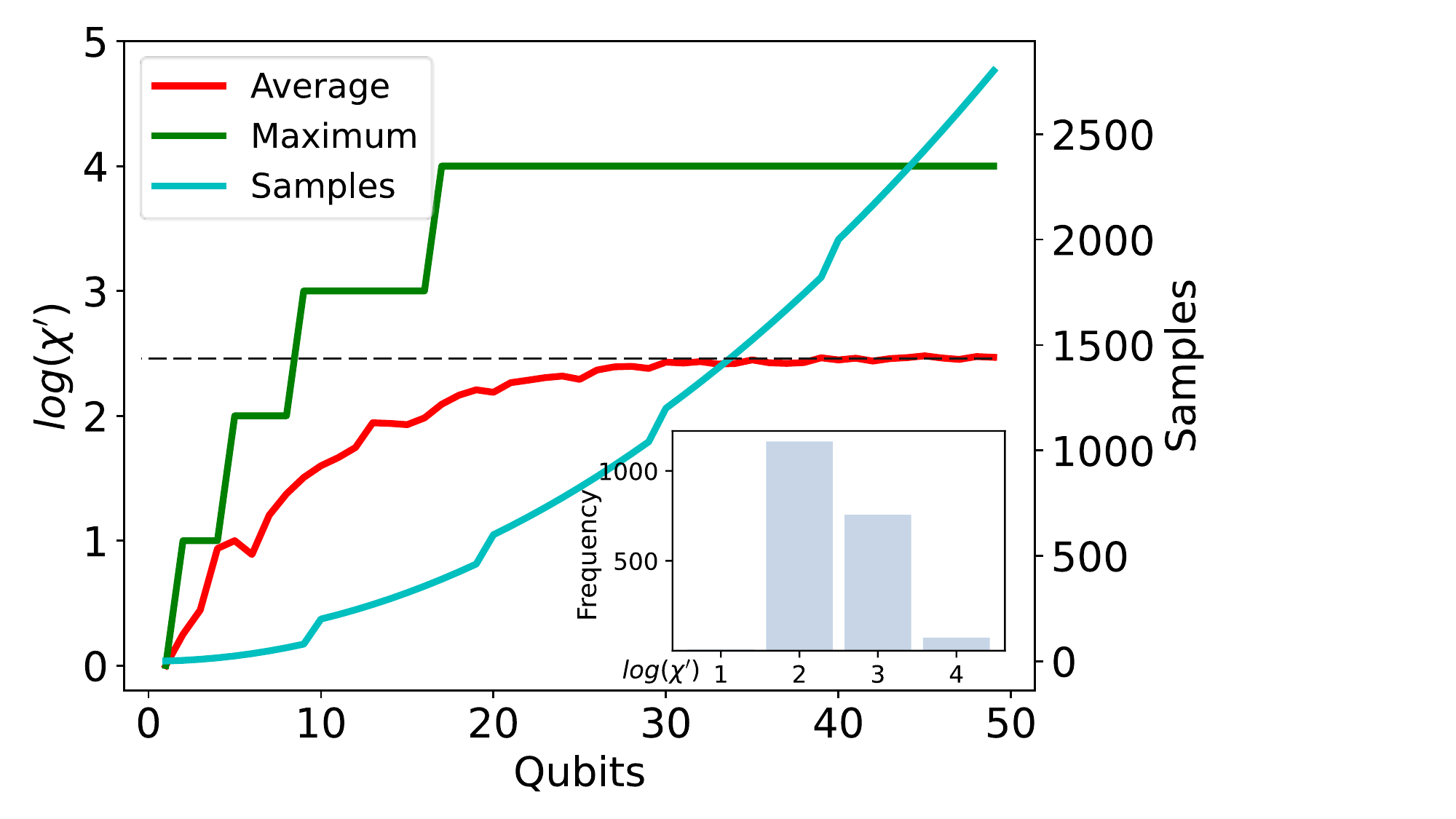}\caption{Average and maximum increase of entanglement after applying a single T-gate on a random Clifford tableau, measured with $log(\chi')$ where $\chi'$ is the maximum bond dimension of the MPS in the stabilizer TN. The average is done over $\sim n^2$ uniformly sampled Clifford circuits. In the inset, the distribution of $log(\chi')$ for $n=40$.}
\label{fig:chi}
\end{figure}
Nonetheless, a gate that entangles $\ket{\psi}$ by a certain amount could in principle become a more entangling gate on $\ket{\nu}$; alternatively, a gate that does not increase stabilizer rank by much could also add a lot of entanglement to $\ket{\nu}$. To show this is not the case, it suffices to look at single-qubit rotations $\mathcal{R}$ for both cases, due to the decomposition we employ. When $\theta$ is not a multiple of $\pi/4$, this rotation is also a good example of an operation that only increases stabilizer rank slightly (Eq. \ref{eq:t_example}). We can bind the growth of $\chi$ in our MPS after applying $\mathcal{R}$ by using Eq. \ref{eq:rule2_nu} (which is equivalent to a CNOT cascade \cite{manskyDecompositionAlgorithmArbitrary2023a}). The worst-case scenario for an MPS is $\chi'=2^4 \chi$, as proven in \hyperref[sec:entangling_power]{Annex \ref{sec:entangling_power}}, although our simulations (Fig. \ref{fig:chi}) show that, on average, it is only $\sim 2^{2.46}$, and that it does not grow as $n\rightarrow\infty$. For other TN structures with more connectivity, this bound can decrease. Regardless, since it is bounded, we can ensure efficient simulation with a low amount of non-Clifford single-qubit rotations. Notice that the worst-case scenario only happens if the circuit applies CNOT gates before $\mathcal{R}$, which were free in the stabilizer TN and stored \textit{potential entanglement} in the basis.\\

\textit{Conclusions and Outlook}: We have presented a new approach to circuit simulation, unifying two different frameworks each with its characterizing resource. We have also shown in which instances it offers an advantage, and identified its free operations for the characterization of a resource theoretic description. In addition, developing the formalism has identified several interesting research directions. First, one could decide to not always use a Clifford gate to update the stabilizer basis of the TN, and apply it directly to $\ket{\nu}$ instead. The criteria used for this decision would directly affect the growth of the resource in the simulation. Also, the storage and retrieval of the \textit{potential} entanglement in $\mathcal{B}(\mathcal{S},\mathcal{D})$ is possible with entangling Clifford gates, but changing the basis also changes $\ket{\nu}$. This means that, during a simulation, we cannot trivially decrease the cost of the next non-stabilizer gate without potentially increasing $\chi$ of the current TN. A thorough study of how to optimally allocate this resource to decrease the cost of the simulation is left for future work.\\

Bounding $\chi$ of the MPS and checking the accuracy of results on states with different amounts of entanglement, magic, or other resources is a strong candidate to characterize the resource $R$, which relates non-trivially to both entanglement and stabilizer rank. In general, being able to relate $\chi$ to a magnitude other than bipartite entanglement also opens up the field of tensor networks to the use of other resources. The evident locality of $\chi$ is an obstacle for which resources can be used in this way, making a link between magic and $\chi$ of our vector $\ket{\nu}$ a specially interesting objective. We can also use stabilizer TNs as a practical tool to bound the simulation hardness of a circuit based on the amount of T-gates it contains, a research that is usually restricted to theoretical approaches \cite{rossOptimalAncillafreeClifford2016}. The method described in this article is available in a Python implementation \footnote{\url{https://github.com/bsc-quantic/stabilizer-TN}} that can simulate any circuit. Thus, one can look into improving the efficiency of the implementation. An obvious candidate is integration with the handling of tableaus as done by STIM \cite{gidneyStimFastStabilizer2021}, the most performant Python approach to stabilizer simulation.\\

\textit{Acknowledgements}: We want to thank Ema Puljak, Berta Casas and Axel Pérez-Obiol for their comments on the manuscript, together with all members of BSC's Quantic group for their suggestions and support.

\bibliography{stab_bibliography}
% {apssamp}
% \bibliography{apssamp}% Produces the bibliography via BibTeX.

\onecolumngrid
\appendix{}

\section*{Annex}
\section{Lemmas and proofs}\label{sec:lemmas}

Here we show and prove the lemmas that generate the \textit{update rules} in the main text. We also include extra comments that make the notation and intuition behind the formalism clearer.

%%%%%%%%% lemma 1

% \newtheorem*{lemma:representation}{Lemma \ref{lemma:representation}}
% \begin{lemma:representation}
\begin{lemma}\label{lemma:representation}
For a given stabilizer basis $\mathcal{B}(S,D)$, any state $\ket{\psi}$ in an n-dimensional Hilbert space $\mathcal{H}^n$ can be described as $\ket{\psi}=\sum_i \nu_i d_{\hat{i}} \ket{\psi_S}$, where $\hat{i}=(i_1 \dots i_n)$, $\nu_i$ are complex coefficients fulfilling $\sum_i |\nu_i|^2 = 1$, and $d_{\hat{i}}= d_1^{i_1}\cdot d_2^{i_2}\cdot\dots\cdot d_n^{i_n}$ with respect to the destabilizer generators $d_i \in D$.
\end{lemma}
% \end{lemma:representation}

While this property underlies the use of stabilizers in error-correction, and thus can be deduced with their formalism, it can also be seen very concisely entirely within the formalism used in this paper. 

\textbf{Proof:} We show that all $d_{\hat{i}}\ket{\psi_S}$ are i) normalized and ii) mutually orthogonal, so they form an orthonormal basis, then that iii) the space they generate is the same dimension as the full n-dimensional Hilbert space.
\begin{itemize}
    \item i) The basis states are normal: $\bra{\psi_S}d_{\hat{i}} d_{\hat{i}}\ket{\psi_S} = \braket{\psi_S|\psi_S} = 1$, using that $\delta_{\hat{i}}\in \mathcal{P}^n$ implies $(\delta_{\hat{i}})^2 = Id$.
    \item ii) The basis states are orthogonal: If we take two different states $d_{\hat{i}}\ket{\psi_S}$, $d_{\hat{j}}\ket{\psi_S}$, then there is a stabilizer generator $d_k$ from D that such that $i_k=1,j_k=0$ or $i_k=0,j_k=1$. Taking the first case without loss of generality, the stabilizer generator $s_k$ anticommutes with $d_{\hat{i}}$ and commutes with $d_{\hat{j}}$, so that
\end{itemize}
    \begin{equation} 
    \bra{\psi_S}d_{\hat{i}} d_{\hat{j}}\ket{\psi_S}=\bra{\psi_S}d_{\hat{i}} d_{\hat{j}} s_k \ket{\psi_S}= - \bra{\psi_S}d_{\hat{i}} s_k d_{\hat{j}} \ket{\psi_S} = - \bra{\psi_S} s_k d_{\hat{i}} d_{\hat{j}} \ket{\psi_S} = -\bra{\psi_S}d_{\hat{i}} d_{\hat{j}}\ket{\psi_S} 
    \end{equation}
    \qquad \: Therefore $\bra{\psi_S}d_{\hat{i}} d_{\hat{j}}\ket{\psi_S} = 0$.
\begin{itemize}   
    \item iii) The basis generates a space of dimension $2^n$: since there are $n$ destabilizers $d_i$, $\hat{i}=(i_1 \dots i_n)$ can take $2^n$ different values, so the basis $\{d_{\hat{i}}\ket{\psi_S}\}_{\hat{i}}$ has that many elements. 
\end{itemize} $\qed$

%%%%%%%%%%%%%%%%%%% Interm 1

Other than the basic structure, we need to understand how arbitrary gates modify $\ket{\psi}$ and $\ket{\nu}$. First, we check how different gates are decomposed into the gates of the basis $\mathcal{B}(\mathcal{S},\mathcal{D})$. Then, we find which operations they correspond to within this formalism. We can see from the definition of the stabilizer basis (Eq. \ref{eq:nu})
% and in Fig. \ref{fig:hilbert_stab} 
that a destabilizer $\delta_{\hat{j}}=d_1^{j_1} \dots d_n^{j_n}$ takes us from one element of the basis to another.
\begin{equation}\label{eq:x_op}
    \delta_{\hat{j}} \ket{\psi} = \delta_{\hat{j}} \sum_{i=0}^{2^n-1} \nu_i d_{\hat{i}} \ket{\psi_\mathcal{S}} = \sum_{i=0}^{2^n-1} \nu_i d_{\hat{i}+\hat{j}} \ket{\psi_\mathcal{S}} = \sum_{i=0}^{2^n-1} \nu_{i+j} d_{\hat{i}} \ket{\psi_\mathcal{S}}.
\end{equation}
On the other hand, the multiplication of a stabilizer $\sigma_{\hat{j}}=s_1^{j_1} \dots s_n^{j_n}$ introduces a sign depending on the element of the basis due to anticommutation. Notice that, because $d_i$ only anticommutes with $s_i$, checking for commutativity is as simple as doing the inner product of their boolean vectors $\hat{d}\cdot \hat{s}$.
\begin{equation}\label{eq:z_op}
    \sigma_{\hat{j}} \ket{\psi} = \sum_{i=0}^{2^n-1} \nu_i \sigma_{\hat{j}} d_{\hat{i}} \ket{\psi_\mathcal{S}} = \sum_{i=0}^{2^n-1} \nu_i (-1)^{\hat{i}\cdot \hat{j}} d_{\hat{i}} \sigma_{\hat{j}} \ket{\psi_\mathcal{S}} = \sum_{i=0}^{2^n-1} (-1)^{\hat{i}\cdot \hat{j}} \nu_i d_{\hat{i}} \ket{\psi_\mathcal{S}}.
\end{equation}
These are equivalent to $X$ and $Z$ operations, respectively, on the computational basis. Therefore, on $\ket{\nu}$ we have:
\begin{equation}\label{eq:notation_xz}
    \delta_{\hat{d}} \ket{\psi} = X_{\hat{d}} \ket{\nu} \quad , \quad \sigma_{\hat{s}} \ket{\psi} = Z_{\hat{s}} \ket{\nu}.
\end{equation}

Since $\mathcal{S}\cup\mathcal{D}$ are a basis for $\mathcal{P}^n$, we can decompose any operator as:
\begin{equation}\label{eq:decomp}
    \mathcal{U} = \sum_i \phi_i \delta_{\hat{d}_i} \sigma_{\hat{s}_i}.
\end{equation}
The previous observations tell us how to apply each factor individually, but any decomposition with more than one term is more complicated. Observe that the difference between the update on $\ket{\psi}$ and on $\ket{\nu}$ is only the changing basis, therefore the transformation to $\ket{\nu}$ must also be a unitary operation. This means that our tensor network representation can use the same tools as with circuit simulation, even if the equivalency is not trivial. The following lemma shows us one useful instance.\\

%%%%%%%%%%% lemma 2

% \newtheorem*{lemma:decomposition}{Lemma \ref{lemma:decomposition}}
% \begin{lemma:decomposition}
\begin{lemma}\label{lemma:decomposition}
For a given stabilizer basis $\mathcal{B}(S,D)$, any unitary that can be decomposed in the form
\begin{equation}
    \mathcal{U} = \phi_1 \delta_{\hat{d}_1} \sigma_{\hat{s}_1} + \phi_2 \delta_{\hat{d}_2} \sigma_{\hat{s}_2},
\end{equation}
is equivalent, in the stabilizer tensor network formalism, to a change of basis with Clifford gates $\delta_{\hat{d}_1}\sigma_{\hat{s}_1}$ followed by a single multi-qubit rotation over the X,Y and Z axes on $\ket{\nu}$:
\begin{equation}\label{eq:lemma2}
    \mathcal{R}_{X_{I_x}Y_{I_y}Z_{I_z}}(2\theta) = \cos(\theta) I - i \sin(\theta) X_{I_x} Y_{I_y} Z_{I_z},
\end{equation}
with $\theta=\arccos{(Re(\phi_1))}$. Using $\circ_h$ 
\footnote{Hadamard multiplication is an element wise multiplication of two tensors $a$ and $b$ of the same shape, such that the entries of the result $c$ follow: $c_{i_1 \dots i_n} = (a \circ_\textit{h} b)_{i_1 \dots i_n} = a_{i_1 \dots i_n} \cdot b_{i_1 \dots i_n}$.}
, the chosen axes $I_x$,$I_y$ and $I_z$ related to $\delta_{\hat{d}_1},\sigma_{\hat{s}_1},\delta_{\hat{d}_2},\sigma_{\hat{s}_2}$ as follows:
\begin{equation}\label{eq:lemma2_Is}
I_y = (\hat{d}_1+\hat{d}_2)\otimes_h(\hat{s}_1+\hat{s}_2) \quad , \quad I_x = (\hat{d}_1+\hat{d}_2) + I_y \quad , \quad I_z = (\hat{s}_1+\hat{s}_2) + I_y
\end{equation}
\end{lemma}
% \end{lemma:decomposition}

\textbf{Proof:}
To make notation a bit easier to read, we refer to operators $\sigma_{\hat{s}_i}$,$\delta_{\hat{d}_j}$ as $\delta_i,\sigma_j$ and use $\delta_i \cdot \sigma_j$ to mean $\hat{s}_i\cdot\hat{d}_j$, except where there might be ambiguity. This helps us keep track of the equations without remembering which operator carries which index. Remember that $\delta \cdot \sigma=1$ when the operators anticommute and $0$ when they commute, and also that any (de)stabilizer is hermitian. It can be checked that unitarity implies the following conditions:
\begin{equation}
\begin{split}
    \mathcal{U}^\dagger \mathcal{U}= I \iff 
    &(\phi_1^* \sigma_1^\dagger \delta_1^\dagger + \phi_2^* \sigma_2^\dagger \delta_2^\dagger)(\phi_1 \delta_1 \sigma_1 + \phi_2 \delta_2 \sigma_2) = \phi_1^* \phi_1 I + \phi_2^* \phi_2 I + \phi_2^* \phi_1 \sigma_2 \delta_2 \delta_1 \sigma_1 + \phi_1^* \phi_2 \sigma_1 \delta_1 \delta_2 \sigma_2 = \\
    &= (\phi_1^* \phi_1 + \phi_2^* \phi_2) I + (\phi_2^*\phi_1 (-1)^{(\sigma_2\cdot\delta_2 + \delta_1\cdot\sigma_2 + \delta_2\cdot\sigma_1 + \sigma_1\cdot\delta_1)} + \phi_1^*\phi_2 ) \sigma_1\delta_1\delta_2\sigma_2 = I \\ 
    \iff & 
    \begin{cases}
        \phi_1 \phi_1^* + \phi_2 \phi_2^* = 1 \\
        \phi_2^*\phi_1 (-1)^{(\delta_1 +\delta_2)\cdot(\sigma_1 + \sigma_2)} = -\phi_1^*\phi_2
    \end{cases}.
\end{split}
\end{equation}
The first condition tells us that we can rewrite the coefficients with trigonometric functions. We can also factorize the complex phase like $\phi_1 = e^{i \varphi_1} \cos(\theta)$ and $\phi_2 = e^{i \varphi_1} e^{i \omega} \sin(\theta)$, and ignore the term $e^{i \varphi_1}$ as a global phase. Substituting this in the second condition:
\begin{equation}\label{eq:lemma2_phase}
\begin{split}
    e^{-i\omega}\sin(\theta)\cos(\theta)(-1)^{(\delta_1 + \delta_2)\cdot(\sigma_1 + \sigma_2)} &= -\cos(\theta) e^{i\omega} \sin(\theta) \rightarrow \\ \rightarrow e^{2i\omega} = (-1)^{(\delta_1 + \delta_2)\cdot(\sigma_1 + \sigma_2)+1} \rightarrow e^{i\omega} &= \pm (-i)^{(\delta_1 + \delta_2)\cdot(\sigma_1 + \sigma_2) + 1}.
\end{split}
\end{equation}
We can also rewrite the unitary as:
\begin{equation}\label{eq:lemma2_uni}
    \mathcal{U} = ( \cos(\theta) I + e^{i\omega} \sin(\theta) \delta_2 \sigma_2 \: \delta_1 \sigma_1 ) \: \delta_1 \sigma_1 
    = ( \cos(\theta) I + e^{i\omega} \sin(\theta) (-1)^{\delta_1\cdot\sigma_2} \delta_2 \delta_1 \: \sigma_2 \sigma_1 ) \: \delta_1 \sigma_1.
\end{equation}
As we have seen in  Eqs. \ref{eq:x_op},\ref{eq:z_op}, we treat a (de)stabilizer operator as a set of Z (X) gates. Therefore, the Pauli operator $\delta_1 \sigma_1$ to the right is strictly a Clifford update (in fact, with only X,Y and Z gates) that we can apply first. It's left to check that the remaining factor behaves like a rotation.
\begin{equation}\label{eq:lemma2_uni2}
    \tilde{\mathcal{U}} =  \cos(\theta) I + e^{i\omega} \sin(\theta) (-1)^{\delta_1\cdot\sigma_2} \delta_2 \delta_1 \: \sigma_2 \sigma_1 
\end{equation}
Similarly, applying $\sigma_{\hat{s}_i}$ ($\delta_{\hat{d}_j}$) is equivalent to the transformation $Z_{\hat{s}_i}$ ($X_{\hat{d}_j}$), and doing two such transformations consecutively is as simple as adding the boolean vectors: $Z_{\hat{s}_j}Z_{\hat{s}_i}=Z_{\hat{s}_j + \hat{s}_i}$. With $I_x$,$I_y$ and $I_z$ defined as above, one can check that
\begin{equation}\label{eq:lemma2_proof_Is}
\begin{split}
I_x + I_y = \hat\delta_1 + \hat\delta_2 &\rightarrow X_{I_x + I_y} = \delta_2 \delta_1\\
I_y + I_z = \hat\sigma_1 + \hat\sigma_2 &\rightarrow Z_{I_y + I_z} = \sigma_2 \sigma_1 \\
I_y = (\hat\delta_1 + \hat\delta_2)\circ_\textit{h}(\hat\sigma_1 + \hat\sigma_2) &\rightarrow |I_y|=\sum_{a\in I_y}a = (\hat\delta_1 + \hat\delta_2)\cdot(\hat\sigma_1 + \hat\sigma_2)
\end{split},
\end{equation}
which is almost the form in Eq.\ref{eq:lemma2}. Notice that $I_x$ has the unique 1s of $\hat{d}_1 + \hat{d}_2$, with $0$s elsewhere, $I_z$ those of $\hat{s}_1 + \hat{s}_2$, whereas $I_y$ has the common $1$s between $\hat{d}_1 + \hat{d}_2$ and $\hat{s}_1 + \hat{s}_2$, with the Hadamard product (element-wise multiplication) enabling the closed form description of $I_y$. Since $X\cdot Z = -iY$, we can rewrite \ref{eq:lemma2} as 
\begin{equation}
    \mathcal{R}_{X_{I_x}Y_{I_y}Z_{I_z}}(2\theta) = \cos(\theta) I + \sin(\theta) (-i)^{|I_y| + 1} X_{I_x+I_y} Z_{I_y+I_z}.
\end{equation}
Putting eq.\ref{eq:lemma2_proof_Is} and eq.\ref{eq:lemma2_phase} together means that the sinus term has the correct phase to be a unitary, so the proposed transformation is indeed a rotation and its coefficients relate to the original unitary as stated. Since the sign can always be changed with the angle of rotation, we are done. $\qed$ \\

The values $v_i$ of the vectors $I_x$ ($I_y$,$I_z$) indicate that we rotate qubit $i$ over $X$ ($Y$,$Z$) if $v_i=1$, and we do nothing if $v_i=0$. Notice that this gate can be implemented with a cascade of CNOT gates on the affected qubits and a single qubit rotation, plus the appropriate basis changes from $X$ to $Y,Z$  \cite{manskyDecompositionAlgorithmArbitrary2023a}. In particular, we implement the rotation on the innermost affected qubit and add CNOT cascades to each side. And example can be seen in Fig. \ref{fig:circ_example}. Lemma \ref{lemma:decomposition} is very useful because the basic $R_X$, $R_Y$ and $R_Z$ gates have this form, so we know how to update with any non-clifford single qubit gate. With the $\ket{\nu}$ notation, lemma \ref{lemma:decomposition} can be summarized with: 
\begin{equation}\label{notation_rot}
    \mathcal{U} \ket{\psi} = (\phi_1 \delta_1 \sigma_1 + \phi_2 \delta_2 \sigma_2 )\ket{\psi} =  \mathcal{R}_{X_{I_x}Y_{I_y}Z_{I_z}}(2\theta)\ket{\nu}
\end{equation}

In the algorithm, we have to check the value of $\delta_1\cdot\sigma_2$ to get the sign of the transformation angle right.\\

We also have a corollary that tells us what the free operations are, which is needed to identify the resource:

%%%%%%%% lemma 2 corollary

\begin{corollary}\label{lemma:free}
In the context of stabilizer simulation as a resource theory, the free operations depend on the basis $\mathcal{B}(S,D)$ and correspond to $\mathcal{U}=\text{cos}(\theta/2) \alpha \delta \sigma + i \text{sin}(\theta/2) d_i s_i \alpha \delta \sigma$, where $d_i\in D$, $s_i \in S$ are generators and $P=\alpha \delta \sigma$ is the decomposition of a Pauli matrix $P$ into $\mathcal{B}(S,D)$.
\end{corollary}

\textbf{Proof:} Since it fits the conditions of lemma \ref{lemma:decomposition}, we apply first the Pauli operator $P$, which consists only of Clifford gates. Then, since $d_i s_i \ket{\psi_S}=d_i\ket{\psi_S}$ is an element of the basis, using \ref{eq:lemma2_Is}, we see that $\cos(\theta/2) + i \sin(\theta/2) d_i s_i$ is equivalent to a single qubit rotation $R_X(-\theta)$ on $\ket{\nu}$, which does not increase its bond dimension. $\qed$ \\

%%%%%%%%%%%%%%%%%%%%%%% Interm 2

For an observable $\mathcal{O}=\alpha \delta_{\hat{n}} \sigma_{\hat{m}} $ we have:
\begin{equation}\label{eq:ev_psi}
\begin{split} 
    \braket{\psi | \mathcal{O} | \psi} = \bra{\psi_\mathcal{S}} \sum_j \nu_{\hat{j}}^* d_{\hat{j}}^* \: \alpha \delta_{\hat{n}} \sigma_{\hat{m}} \sum_i \nu_{\hat{i}} d_{\hat{i}} \ket{\psi_\mathcal{S}} &= \sum_{i,j} \alpha \nu_{\hat{j}}^* \nu_{\hat{i}} (-1)^{{\hat{m}}\cdot {\hat{i}}} \braket{\psi_{\mathcal{S}} |d_{\hat{j}} d_{\hat{n}} d_{\hat{i}} | \psi_{\mathcal{S}}} = \\ = \sum_{i,j} \alpha \nu_{\hat{j}}^* \nu_{\hat{i}} (-1)^{{\hat{m}}\cdot {\hat{i}}} \delta'_{\hat{j},\hat{i}+\hat{n}} &= \alpha \sum_i (-1)^{{\hat{m}}\cdot {\hat{i}}} \nu^*_{\hat{i}+\hat{n}} \nu_{\hat{i}},
\end{split}
\end{equation}
where $\delta'$ is a Kronecker delta. This is much simpler on $\ket{\nu}$:
\begin{equation}\label{eq:ev_nu}
    \braket{\psi | \mathcal{O} | \psi} = \alpha \braket{\nu | X_{\hat{n}} Z_{\hat{m}} |\nu}.
\end{equation}
Notice that the $\alpha$ phase comes from forcing a specific decomposition on $\mathcal{B}(\mathcal{S},\mathcal{D})$, which might mean we have to write $XZ=-iY$ instead of $Y$ directly; it does not mean we allow non-physical observables, that is, with non-real expected values. A measurement of this observable $\mathcal{O}$ projects the state $\ket{\psi}$:
\begin{equation}\label{eq:projection}
    \ket{\psi} \rightarrow \frac{I \pm \mathcal{O}}{2} \ket{\psi}.
\end{equation}
The sign is $+$ ($-$) when projecting to the positive $\ket{\psi_+}$ (negative $\ket{\psi_-}$) eigenstate. We can calculate $\braket{\mathcal{O}}$ with Eq. \ref{eq:ev_nu} and randomly decide the output with probability $p=\frac{1+\braket{\mathcal{O}}}{2}$ for $\ket{\psi_+}$ and $1-p=\frac{1-\braket{\mathcal{O}}}{2}$ for $\ket{\psi_-}$. Then the following lemma shows how to update the coefficients.\\

%%%%%%%%%%% lemma 3

% \newtheorem*{lemma:measurement}{Lemma \ref{lemma:measurement}}
% \begin{lemma:measurement}
\begin{lemma}\label{lemma:measurement}
For a given stabilizer basis $\mathcal{B}(S,D)$ and an observable $\mathcal{O}$ that decomposes as
\begin{equation}
	\mathcal{O} = \alpha \delta_{\hat{n}} \sigma_{\hat{m}},
\end{equation}
the projection $\braket{I\pm\mathcal{O}}{2}$ onto the positive (negative) eigenstate is equivalent to the following non-unitary operation on $\ket{\nu}$:
\begin{equation}\label{eq:lemma3_rot_annex}
    P_k \cdot \tilde{\mathcal{R}}_{X_{I_x}Y_{I_y}Z_{I_z}} = P_k \cdot \left( \frac{1}{\sqrt{2}} I \pm \frac{\alpha (-i)^{|I_y|}}{\sqrt{2}} X_{I_x} Y_{I_y} Z_{I_z} \right),
\end{equation}
where $k$ is the position of the first $1$ in $\hat{n}$, $P_k$ is the projector $\ket{0}\bra{0}$ on qubit $k$, and the choice of rotation axes is given by $\delta_{\hat{n}},\sigma_{\hat{m}}$ as
\begin{equation}\label{eq:lemma3_Is_annex}
I_y = \hat{n}\circ_h \hat{m} \quad , \quad I_x = \hat{n} + I_y \quad , \quad I_z = \hat{m} + I_y
\end{equation}
The resulting state $\ket{\psi'}$ is a valid quantum state when renormalized as $\sqrt{\frac{2}{1\pm\braket{\psi | \mathcal{O} | \psi}}}\ket{\psi'}$.
\end{lemma}
% \end{lemma:measurement}

\textbf{Proof:} We can expand the projection of $\frac{I\pm\mathcal{O}}{2}$ into:
\begin{equation}\label{eq:lemma3_proj}
    \frac{I\pm\mathcal{O}}{2} \ket{\psi} = \frac{1}{2} \sum_{\hat{i}} (I \pm \alpha \delta_{\hat{n}} \sigma_{\hat{m}} ) \nu_{\hat{i}} d_{\hat{i}} \ket{\psi_{\mathcal{S}}} = \frac{1}{2} \sum_{\hat{i}} ( \nu_{\hat{i}} \pm \alpha (-1)^{\hat{i}\cdot \hat{m}} \nu_{\hat{i}+\hat{n}} ) d_{\hat{i}} \ket{\psi_{\mathcal{S}}}.
\end{equation}
Then we must consider the update to the stabilizer basis. When $\hat{n}=0$, we are projecting onto a stabilizer of $\ket{\psi_\mathcal{S}}$ and there is no update to $\mathcal{B}(\mathcal{S},\mathcal{D})$. In this case we are directly left with:
\begin{equation}\label{eq:lemma3_proj_n0}
    \frac{I\pm\mathcal{O}}{2} \ket{\psi} = \frac{1}{2} \sum_{\hat{i}} \nu_{\hat{i}} (1 \pm \alpha (-1)^{\hat{i}\cdot \hat{m}} ) d_{\hat{i}} \ket{\psi_{\mathcal{S}}} \quad \text{if } \hat{n}=0.
\end{equation}
In the case $\delta_{\hat{n}}\neq 0$, it was shown in \cite{yoderGeneralizationStabilizerFormalism} how to update the basis in terms similar to our Eq. \ref{eq:lemma3_proj}. Adapting those results to our notation for $\ket{\nu}$, we get the new basis $\mathcal{B'}(\mathcal{S}',\mathcal{D}')$ and :
\begin{equation}\label{eq:yoder_meas_1}
    \frac{I \pm \mathcal{O}}{2} \ket{\psi} = \sum_{\hat{i}} \left[ \frac{1}{2} (\pm \alpha (-1)^{\hat{i}\cdot \hat{m}} )^{\hat{i}_k} \nu_{\hat{i}} \right] \: d_{\hat{i}+\hat{i}_k \cdot \hat{n}}' \ket{\psi_{\mathcal{S}'}} \quad \text{if } \hat{n} \neq 0,
\end{equation}
where $k$ is the position of the first $1$ in $\hat{n}$ and $\hat{i}_k$ the $k^{th}$ element of $\hat{i}$. Notice that $\hat{i}_k = 0$ implies $(\hat{i}+\hat{n})_k = 1$, that is, $d_{\hat{i}+\hat{i}_k\cdot \hat{n}}=d_{\hat{i}}$ and $d_{(\hat{i}+\hat{n})+(\hat{i}+\hat{n})_k\cdot \hat{n}}=d_{(\hat{i}+\hat{n})+\hat{n}}$, so the coefficient $\frac{1}{2} (\alpha (-1)^{\hat{i}\cdot \hat{m}} )^{\hat{i}_k} \nu_{\hat{i}}$ stays in $d_{\hat{i}}\ket{\psi_{\mathcal{S}}}$ and $\frac{1}{2} (\alpha (-1)^{(\hat{i}+\hat{n})\cdot \hat{m}} )^{(\hat{i}+\hat{n})_k} \nu_{\hat{i}+\hat{n}}$ moves to $d_{(\hat{i}+\hat{n})+\hat{n}}\ket{\psi_{\mathcal{S}}}=d_{\hat{i}}\ket{\psi_{\mathcal{S}}}$, leaving $\nu_{\hat{i}+\hat{n}}$ empty; for $\hat{i}_k=1$ both coefficients concentrate on $\nu_{\hat{i}+\hat{n}}$ and leave $\nu_{\hat{i}}$ empty instead: the measurement halves the non-zero coefficients whenever $\hat{n}\neq 0$. This means we can rewrite the above as:
\begin{equation}\label{eq:meas_sup}
    \frac{1\pm\mathcal{O}}{2} \ket{\psi} = \sum_{\hat{i}} \delta_{\hat{i}_k,0} \left( \frac{1}{\sqrt{2}} \nu_{\hat{i}} \pm \frac{\alpha (-1)^{\hat{i}\cdot \hat{m}}}{\sqrt{2}}\nu_{\hat{i}+\hat{n}} \right) d_{\hat{i}} \ket{\psi_{\mathcal{S}}} \quad \text{if } n \neq 0.
\end{equation}
where the Kronecker delta filters the non-zero coefficients. Defining $\hat{i}_k\equiv0$ when $\hat{n}=0$, we see the only difference between Eq. \ref{eq:lemma3_proj_n0} and Eq. \ref{eq:meas_sup} is a factor of $\sqrt{2}$. Since we have to normalize at the end anyway, we can rejoin both cases and proceed with Eq. \ref{eq:meas_sup}. In terms of $\ket{\nu}$, we can prepare the superposition on all states, which looks almost like a rotation:
\begin{equation}\label{eq:meas_rot1}
    \tilde{\mathcal{R}}\ket{\nu} = \tilde{\mathcal{R}}\sum_i \nu_{\hat{i}} = \sum_i \frac{1}{\sqrt{2}} \nu_{\hat{i}} \pm \frac{\alpha (-1)^{\hat{i}\cdot \hat{m}}}{\sqrt{2}} \: \nu_{\hat{i}+\hat{n}},
\end{equation}
and remove the duplicate coefficients \textit{afterwards} with a projection on the $\ket{0}$ state of qubit $k$:
\begin{equation}\label{eq:meas_proj}
	P_k = I_0 \otimes \dots I_{k-1} \otimes \begin{pmatrix} 1 & 0 \\ 0 & 0 \end{pmatrix} \otimes I_{k+1} \dots \otimes I_n \equiv \ket{0}\bra{0}_k,
\end{equation} 
so that $\frac{I+\mathcal{O}}{2} \ket{\nu} = P_k \: \tilde{\mathcal{R}} \ket{\nu}$. This transformation is similar to the rotation in lemma \ref{lemma:decomposition}, but removing the $i$ phase. We can reuse the reasoning there to find:
\begin{equation}\label{eq:meas_rot2}
	\tilde{\mathcal{R}}_{X_{I_x}Y_{I_y}Z_{I_z}} = \frac{1}{\sqrt{2}} I \pm \frac{\alpha (-i)^{|I_y|}}{\sqrt{2}} X_{I_x} Y_{I_y} Z_{I_z},
\end{equation}
where $I_x,I_y,I_z$ are related to $\delta_{\hat{n}},\sigma_{\hat{m}}$ exactly as in Eq. \ref{eq:lemma3_Is_annex}. Now, $I_x$ has the unique 1s of $\hat{n}$, with $0$s elsewhere, $I_z$ those of $\hat{m}$, whereas $I_y$ has the common $1$s between $\hat{n}$ and $\hat{n}$. Although it is not a unitary operation, it's equivalent to a projection, so the output is not normalized but is otherwise a valid state. To find the normalization term, we reuse Eq. \ref{eq:ev_psi}:
\begin{equation}\label{lemma3_ev}
    \braket{\psi|\mathcal{O}|\psi} = \alpha \sum_{\hat{i}} (-1)^{\hat{i}\hat{m}}\nu^*_{\hat{i}+\hat{n}} \nu_{\hat{i}} = \alpha^* \sum_{\hat{i}} (-1)^{\hat{i}\hat{m}} \nu^*_{\hat{i}} \nu_{\hat{i}+\hat{n}}.
\end{equation}
The second equality is a consequence of $\braket{\psi|\mathcal{O}|\psi}$ being real. Now:
\begin{equation}\label{eq:lemma3_norm_sq}
\begin{split}
    & \qquad \qquad \qquad \qquad \qquad \mathcal{N}^2 = \left( \bra{\psi_{\mathcal{S}}} \frac{1\pm\mathcal{O}^\dagger}{2} \right) \left(\frac{1\pm\mathcal{O}}{2} \ket{\psi_{\mathcal{S}}} \right) = \\
    &= \sum_{\hat{i},\hat{j}} \bra{\psi_{\mathcal{S}}} d_{\hat{j}} \left( \frac{1}{\sqrt{2}}\nu_{\hat{j}}^*\pm \frac{\alpha^*(-1)^{\hat{j}\cdot \hat{m}}}{\sqrt{2}}\nu_{\hat{j}+\hat{n}}^* \right)\delta_{\hat{j}_k,0}\delta_{\hat{i}_k,0}\left( \frac{1}{\sqrt{2}}\nu_{\hat{i}}\pm \frac{\alpha^*(-1)^{\hat{i}\cdot \hat{m}}}{\sqrt{2}}\nu_{\hat{i}+\hat{n}} \right) d_{\hat{i}} \ket{\psi_{\mathcal{S}}} = \\
    &= \sum_{\hat{i},\hat{j}} \delta_{\hat{j}_k,0}\delta_{\hat{i}_k,0} \left( \frac{1}{\sqrt{2}}\nu_{\hat{j}}^*\pm \frac{\alpha^*(-1)^{\hat{j}\cdot \hat{m}}}{\sqrt{2}}\nu_{\hat{j}+\hat{n}}^* \right) \left(\frac{1}{\sqrt{2}} \nu_{\hat{i}}\pm \frac{\alpha^*(-1)^{\hat{i}\cdot \hat{m}}}{\sqrt{2}}\nu_{\hat{i}+\hat{n}} \right) \braket{\psi_{\mathcal{S}} | d_{\hat{j}} d_{\hat{i}} | \psi_{\mathcal{S}}} = \\
    &= \sum_{\hat{i},\hat{j}} \delta_{\hat{j}_k,0}\delta_{\hat{i}_k,0} \left( \frac{1}{2}\nu_{\hat{j}}^*\nu_{\hat{i}} \pm \frac{\alpha^*(-1)^{\hat{j}\cdot \hat{m}}}{2}\nu_{\hat{j}+\hat{n}}^*\nu_{\hat{i}} \pm \frac{\alpha(-1)^{\hat{i}\cdot \hat{m}}}{2}\nu_{\hat{j}}^* \nu_{\hat{i}+\hat{n}} + \frac{|\alpha|^2}{2}v_{\hat{j}+\hat{n}}^*v_{\hat{i}+\hat{n}} \right) \delta_{\hat{i},\hat{j}} = \\
    & \quad = \sum_{\hat{i}} \delta_{\hat{i}_k,0} \left( \frac{1}{2}|\nu_{\hat{i}}|^2 + \frac{|\alpha|^2}{2}v_{\hat{i}+\hat{n}}^*v_{\hat{i}+\hat{n}} \pm \frac{\alpha^*(-1)^{\hat{i}\cdot \hat{m}}}{2}\nu_{\hat{i}+\hat{n}}^*\nu_{\hat{i}} \pm \frac{\alpha(-1)^{\hat{i}\cdot \hat{m}}}{2}\nu_{\hat{i}}^* \nu_{\hat{i}+\hat{n}} \right) = \\
    & \qquad \qquad \qquad \quad = \frac{1}{2} \sum_{\hat{i}} \delta_{\hat{i}_k,0} |\nu_{\hat{i}}|^2 + \frac{1}{2} \sum_{\hat{i}} \delta_{\hat{i}_k,0}|\nu_{\hat{i}+\hat{n}}|^2 \pm \sum_{\hat{i}} \delta_{\hat{i}_k,0} \braket{\psi|\mathcal{O}|\psi} = \\ 
    & \qquad \qquad \qquad  = \frac{1}{2} \sum_{\hat{i}} (\delta_{\hat{i}_k,0} + \delta_{(\hat{i}+\hat{n})_k,0})|\nu_{i}|^2 \pm \frac{1}{2} \braket{\psi|\mathcal{O}|\psi} = \frac{1\pm \braket{\psi|\mathcal{O}|\psi}}{2}.
\end{split}
\end{equation}
We used that $\hat{i}_k=0\leftrightarrow (\hat{i}+\hat{n})_k=1$ again, so the sum of $(\delta_{\hat{i}_k,0} + \delta_{(\hat{i}+\hat{n})_k,0})$ is always $1$. Because $(\hat{i}+\hat{n})+\hat{n}=\hat{i}$, each contribution to the sum appears twice, so the Kronecker delta selects one of each pair and is thus equivalent to the factor $\frac{1}{2}$ in front of the expected value. This equation proves we have a valid quantum state in all cases with a renormalization term of:
\begin{equation}\label{eq:lemma3_norm}
    \mathcal{N} = \sqrt{\frac{1 \pm \braket{\psi|\mathcal{O}|\psi}}{2}}.
\end{equation} 
$\qed$

We can implement the rotation we found with a CNOT cascade similarly to Eq. \ref{eq:lemma2}, but instead of a central $R_X$ rotation we need the following one-qubit (non-unitary) operation:
\begin{equation}\label{eq:meas_central}
	\tilde{\mathcal{R}} = \frac{1}{\sqrt{2}} \begin{pmatrix} 1 & \pm \alpha (-i)^{|I_y|} \\ \pm \alpha (-i)^{|I_y|} & 1 \end{pmatrix}.
\end{equation}

%%%%%%%%%%%%%%%%%%%%%% End of lemmas

\section{Entangling power of an arbitrary gate}\label{sec:entangling_power}

\begin{figure*}[!ht]
\centering
\includegraphics[scale=0.6]{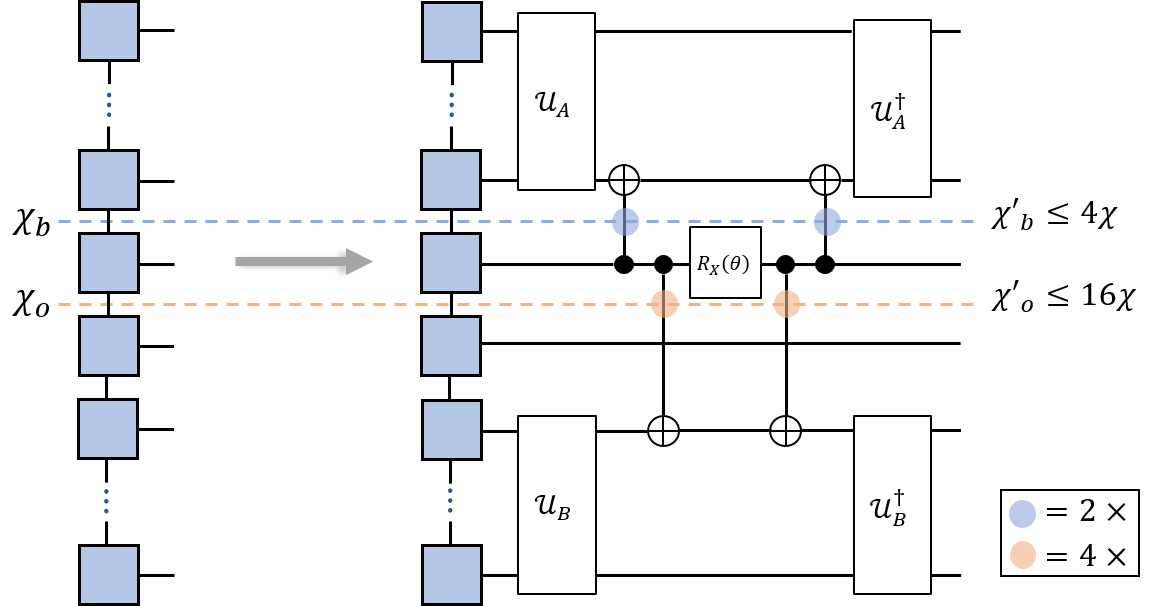}
\caption{Possible decomposition of the rotation \ref{eq:rule2_nu} showcasing how the gates affect $\chi$ for different bonds. All gates that can be  grouped into a unitary on partition A (B) are irrelevant. The bond in blue starts with $\chi_b$ and each CNOT can increase it by at most a factor of $2$, reaching $\chi_b'\leq 4\chi_b$ after the transformation. For the orange bond, starting with $\chi_o$, the implementation of a CNOT across far away qubits on an MPS requires that we apply a SWAP gate at each line crossing, which increases $\chi$ by at most $4$, so that at the end we have $\chi_o'\leq 16\chi_o$. This argument is independent of the initial bonds.
\label{fig:entang_example}}
\end{figure*}

There is not a unique way to define the entangling power of a gate. There are many notions of multipartite entanglement that one can use, and the entanglement of the final state depends on the initial state. This means we have to look at an average over all possible states or a worst/best case, which can also complicate the calculation of the chosen metric. One of the first proposals \cite{zanardiEntanglingPowerQuantum2000} used the average linear entropy over all possible product states $\ket{\psi_A}\otimes\ket{\phi_B}$ on a preset bipartition $A,B$ of the whole space:
\begin{equation}\label{eq:ent_power}
    e(U) \coloneqq \overline{E(U\ket{\psi_A}\otimes\ket{\phi_B})}^{\psi_A,\phi_B}.
\end{equation}
This approach has appeared specially useful when studying entangling power on mixed states \cite{guanEntanglingPowerTwoqubit2014}. Others are based on unitary evolution and focus on the norm of an infinitesimal transformation, such as \cite{marienEntanglementRatesStability2016,eisertEntanglingPowerQuantum2021a}. Metrics other than entanglement have also been used, such as quantum discord \cite{galveDiscordingPowerQuantum2013}. Since we deal with an MPS, we focus on the bipartite entanglement case, and instead of the average, we find the maximum bond dimension $\chi'$ needed in our MPS after applying a gate on an MPS that had maximum bond dimension $\chi$. We use the Schmidt decomposition of unitaries \cite{nielsenQuantumDynamicsPhysical2003}, which has been used previously to characterize arbitrary gates \cite{jonnadulaEntanglementMeasuresBipartite2020} and tells us we can decompose any unitary as 
\begin{equation}\label{eq:cnot_chi}
    \mathcal{U} = \sum_{i=1}^k s_i A_i \otimes B_i,
\end{equation}
where $s_i\geq 0$, $\sum_i |s_i|^2 = 1$ and $A_i, B_i$ are an orthogonal operator basis \cite{balakrishnanOperatorSchmidtDecompositionGeometrical2011}. When applying the rotation $\mathcal{R}$ in Eq. \ref{eq:rule2_nu}, we can focus on any arbitrary bond of our MPS by decomposing it as in Fig. \ref{fig:entang_example}. This way, we only need to check how the gates that cross the chosen bond affect $\chi$.\\

With the Schmidt decomposition \cite{Nielsen_Chuang_2010} of the initial state $\ket{\psi}$ as:
\begin{equation}\label{eq:state_chi}
    \ket{\psi} = \sum_{i=1}^\chi \lambda_i \ket{\psi_A^i} \otimes \ket{\psi_B^i},
\end{equation}
limited to rank $\chi$, applying a two qubit gate with Schmidt number $k$ means we get that $\chi'$ is at most:
\begin{equation}\label{eq:cnot_state_chi}
    \mathcal{U}_k\ket{\psi} = \sum_{j=1}^k \sum_{i=1}^\chi (s_j \lambda_i) A_j \ket{\psi_A^i} \otimes B_j \ket{\psi_B^i}. = \sum_{i=1}^{k\chi} \tilde{\lambda}_i \ket{\phi_A^i} \otimes \ket{\phi_B^i}
\end{equation}
The final form is still a valid Schmidt decomposition thanks to the orthonormality of $A_i, B_i$ and $\sum_is_k^2=1$. A CNOT gate has Schmidt number $k=2$ \cite{balakrishnanOperatorSchmidtDecompositionGeometrical2011}, so the set of two CNOTs that are applied to a particular bond can increase at most $\chi'\leq4\chi$. Counter to intuition, the worst-case scenario when using an MPS is not an update that affects all qubits, but one that affects qubits that are far apart: a CNOT gate over tensors that are not neighbours is implemented with SWAPs on our MPS. A SWAP gate has Schmidt rank 4, so Eq. \ref{eq:cnot_state_chi} gives the bound $\chi'\leq16\chi$ instead, as stated in the main text and fitting the simulations in \ref{fig:chi}. Since this maximum is a consequence of SWAP gates, TN geometries other than MPS that adapt to the connectivity of the simulated circuit can reduce the bound to $4\chi$; this entails a bigger complexity in the TN contraction, as is the case in general for higher dimensional networks.\\

\section{Stabilizer TN update example}

\begin{figure*}[!ht]
\centering
\includegraphics[scale=0.6]{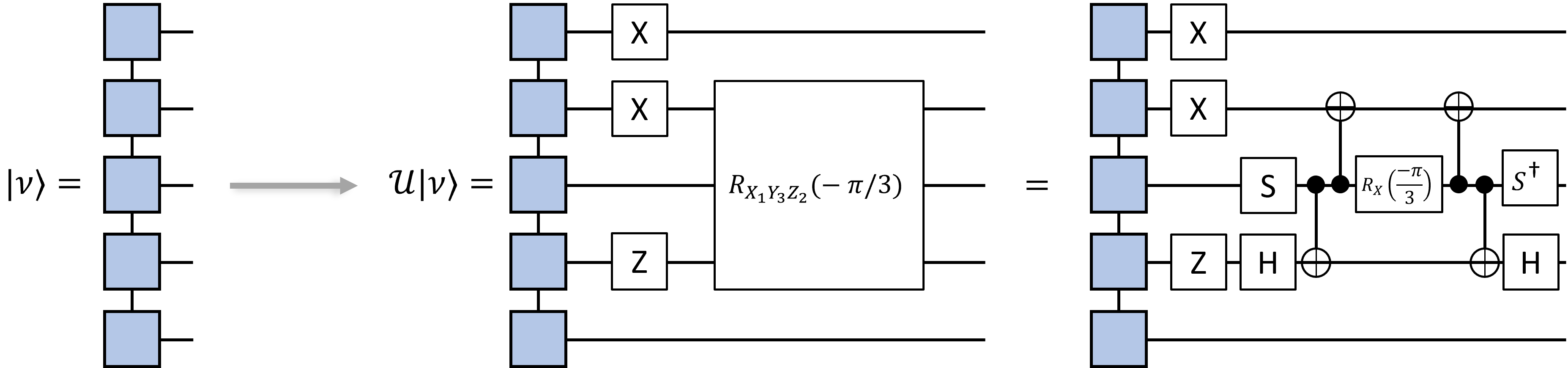}
\caption{Example of coefficient update for the unitary described in Eq. \ref{eq:example_unitary}. Horizontal lines are "qubit" sites in the traditional MPS tensor network representation of a quantum state. Gates are applied from left to right. The resulting TN of this example is more entangled than the initial $\ket{\nu}$.
\label{fig:circ_example}}
\end{figure*}

It is useful to illustrate an example of a coefficient update with a non-Clifford gate in terms of $\ket{\nu}$. We take an arbitrary basis $\mathcal{B}(\mathcal{S},\mathcal{D})$ for $5$ qubits and a unitary that decomposes as
\begin{equation}\label{eq:example_unitary}
    \mathcal{U} = \frac{\sqrt{3}}{2}\delta_{\hat{d}_1}\sigma_{\hat{s}_1} + \frac{1}{2}\delta_{\hat{d}_2}\sigma_{\hat{s}_2} = \left( \cos(\pi/6) + \sin(\pi/6)\delta_{\hat{d}_2}\sigma_{\hat{s}_2}\delta_{\hat{d}_1}\sigma_{\hat{s}_1} \right) \delta_{\hat{d}_1}\sigma_{\hat{s}_1} = \left( \cos(\pi/6) - \sin(\pi/6)(\delta_{\hat{d}_2}\delta_{\hat{d}_1})(\sigma_{\hat{s}_2}\sigma_{\hat{s}_1}) \right) \delta_{\hat{d}_1}\sigma_{\hat{s}_1}.
\end{equation}
Let us assume that the vector representation of $\delta,\sigma$ is:
\begin{equation}\label{eq:example_operators}
\begin{split}
    \left.\begin{array}{c} \hat{d}_1 = (1,1,0,0,0) \\ \hat{d}_2 = (1,0,0,1,0) \end{array}\right\rbrace & \rightarrow \hat{d}_2\hat{d}_1 = (0,1,0,1,0) \\
    \left.\begin{array}{c} \hat{s}_1 = (0,0,0,1,0) \\ \hat{s}_2 = (0,0,1,0,0) \end{array}\right\rbrace & \rightarrow \hat{s}_2 \hat{s}_1 = (0,0,1,1,0)
    % \delta_1 = (1,1,0,0,0)
    % \delta_2 = (1,0,0,1,0) 
    % \sigma_1 = (0,0,0,1,0)
    % \sigma_2 = (0,0,1,0,0)
\end{split},
\end{equation}
which also implies $\hat{d}_1 \cdot \hat{s}_2 = 1$. Then using Eqs. \ref{eq:x_op},\ref{eq:z_op} and the $\ket{\nu}$ notation we can rewrite 
\begin{equation}\label{eq:example_rotation}
    \mathcal{U}\ket{\nu} = \left(\cos(\pi/6)-\sin(\pi/6)(X_1X_3)(Z_2Z_3)\right) X_0X_1Z_3 \ket{\nu} = \left(\cos(\pi/6)+i\sin(\pi/6)X_1Y_3Z_2\right) X_0X_1Z_3 \ket{\nu}
\end{equation}
We see that this fits Eq. \ref{eq:rule2_nu}. Then the coefficients can be updated with the corresponding multiqubit rotation, which we show in Fig. \ref{fig:circ_example}. We also show the decomposition that we have used in our Python implementation, which uses two cascades centred on the middle qubit instead of a single CNOT cascade.\\

Most circuit simulations compile an input gate set into a specific set of gates, since practical realizations of quantum computers are similarly bound by a limited set of native gates. Our simulation approach can handle any circuit with a $\{CNOT,R_X,R_Y,R_Z\}$ decomposition, so it's compatible with most circuits despite the limitations of lemma \ref{lemma:decomposition}. Other characterizations are possible and still compatible with the stabilizer TN framework, but the implementation of unitaries of arbitrary decomposition is left for future work.

\section{Tableau update rules}\label{sec:tableau_rules}

Our formalism relies on the update rules for the original tableau:

\begin{equation}
% \left(\begin{array}{cc|*{3}{c}}
\left(\begin{array}{ccc|ccc|c}
x_{1,1}  & \cdots &  x_{1,n}  & z_{1,1} & \cdots & x_{1,n} & r_1    \\
\vdots  & \ddots & \vdots   & \vdots & \ddots & \vdots & \vdots \\
x_{n,1}  & \cdots & x_{n,n}   & z_{n,1} & \cdots & z_{n,n} & r_n    \\
    \hline
x_{n+1,1} & \cdots & x_{n+1,n} & z_{n+1,1} & \cdots & x_{n+1,n} & r_{n+1} \\
\vdots     & \ddots & \vdots     & \vdots     & \ddots & \vdots     & \vdots    \\
x_{2n,1}  & \cdots & x_{2n,n}  & z_{2n,1}  & \cdots & z_{2n,n}  & r_{2n}    \\
\end{array}\right).
\end{equation}

This section follows \cite{aaronsonImprovedSimulationStabilizer2004} exactly. It is included to give a self-contained explanation of our formalism's update rules, since the original update rules are also used. Considering that a measurement over $X$ or $Y$ basis can be set to a Clifford operator followed by a $Z$ basis measurement, the basic operations we need are, using always base 2:
\begin{enumerate}
        \item \textbf{CNOT operator with control qubit $a$ and target qubit $b$:} For every row $i$, update entries $\textit{x}_{ib}$, $\textit{z}_{ia}$, $\textit{r}_{i}$ as
        \begin{equation}
            r_i := r_i \oplus x_{ia} z_{ib}(x_{ib}\oplus z_{ia}\oplus 1) \quad ; \quad x_{ib}:=x_{ib}\oplus x_{ia} \quad ; \quad z_{ia}:=z_{ia}\oplus z_{ib}.
        \end{equation}
        \item \textbf{Hadamard operator on qubit $a$:}  For every row $i$, update entries $\textit{x}_{ia}$, $\textit{z}_{ia}$, $\textit{r}_{i}$ as
        \begin{equation}
            r_i := r_i \oplus x_{ia} z_{ia} \quad ; \quad x_{ia}:=z_{ia} \quad ; \quad z_{ia}:=x_{ia}.
        \end{equation}
        \item \textbf{Phase operator on qubit $a$:} For every row $i$, update entries  $\textit{z}_{ia}$, $\textit{r}_{i}$ as
        \begin{equation}
            r_i := r_i \oplus x_{ia} z_{ia} \: ; \: z_{ia}:=z_{ia}\oplus x_{ia}.
        \end{equation}
        \item \textbf{Measurement over $Z$ basis on qubit $a$:} The update is different whether the measurement commutes with the current stabilizers. If it does, we do not need to update the tableau. Then, for each row $i\in\{1\dots 2n\}$, we do the operation \textbf{rowsum(i,t)} on an auxiliary row $t$ that starts with all zeroes, and the phase of $t$ tells us if we measure 0 or 1. Otherwise, it must anticommute, so the outcome $r$ is random with equal probability, but we must change the tableau. To do so, we choose the row $i$ of one of the anticommuting stabilizers $H$ (those with $x_{ia}=1$), do the operation \textbf{rowsum(i,h)} for all $h\in H\setminus \{i\}$, and finally we add the observable $Z_a$ to the list of stabilizers at $i$ and store the former stabilizer $i$ as a destabilizer at row $i-n$.
\end{enumerate}
While not technically an operation that appears in the circuit, we also need to define what this "\textbf{rowsum}" operation does:
\begin{enumerate}[resume]
        \item \textbf{rowsum(a,b):} Sets generator $a$ to $a+b$, that is, $x_{aj} = x_{aj}\oplus x_{bj}$ and $z_{aj} = z_{aj}\oplus z_{bj}$ for all $j\in\{1\dots n\}$, while properly changing its phase too. To do so, we need a function $g(x_1, z_1, x_2, z_2)$ that returns the exponent of the phase we get from multiplying $x_1z_1 \cdot x_2z_2$. That is:
        \begin{equation}
            \begin{split}
                x_1 = z_1 = 0  &\rightarrow g=0 \\
                x_1 = z_1 = 1 &\rightarrow g = z_2 - x_2 \\
                x_1 = 1, z_1 = 0 &\rightarrow g = z_2 (2x_2 - 1) \\
                x_1 = 0,z_1 = 1 &\rightarrow g = x_2 (1 - 2z_2)
            \end{split}.
        \end{equation}
        Then the phase $r_a$ is:
        \begin{equation}
        \begin{split}
        r_a = 0 \text{  if  } 2r_a + 2r_b + \sum_{j=1}^n g(x_{bj},z_{bj},x_{aj},z_{aj} \equiv 0 \: (\text{mod } 4) \\
        r_a = 1 \text{  if  } 2r_a + 2r_b + \sum_{j=1}^n g(x_{bj},z_{bj},x_{aj},z_{aj} \equiv 2 \: (\text{mod } 4) \\
        \end{split}.
        \end{equation}
\end{enumerate}

\end{document}